\documentclass[a4paper,11pt]{article}
\usepackage{epsfig}
\advance \topmargin by -\headheight
\advance \topmargin by -\headsep
\evensidemargin \oddsidemargin
\advance\hoffset by -3mm  

\title{Problems in realization of large-scale ensemble silicon-based NMR quantum computers}
\author{A.A.Kokin, K.A.Valiev}
\date{}

\begin{document}
\maketitle

\par
\par
\par
\thanks{Institute of Physics and Technology of Russian Academy of Sciences, 34, Nakhimovskii pr., 117218 Moscow, Russia}
\par
\begin{abstract}
Problems in realization of silicon-based solid-state NMR quantum
computer with ensemble addressing to qubits are considered. It is
presented the extension of Kane's scheme to ensemble approach
version with strip gates. For the initialization of nuclear quantum
states it is proposed to use the solid-state effect in ENDOR
technique whereby the nuclear spins can be practically fully
polarized or, that is the same, indirect cooled to the spin
temperature less than $\sim 1\,\mathrm{mK}$. It is suggested the possible planar
silicon topology of such ensemble quantum computer and shown that the
measurement with standard NMR methods signal of $L \sim 10^{3}$ qubit system
may be achieved for a number of ensemble components $N \geq 10^{5}$. As another
variant of ensemble silicon quantum computer the gateless
architecture of cellular-automaton is also considered. The
decoherence of quantum states in the ensemble quantum computers and
ways of its suppression is also discussed.\par
\end{abstract}

\section*{Introduction}
\par
Atomic nuclei with a spin quantum number $I = 1/2$ seem to be the 
{\it natural candidates }for qubits --- two-level quantum elements in quantum 
computers. The early approach to NMR quantum computers was suggested 
in 1997 \cite{1,2} and then confirmed in experiments \cite{3,4}. In this
approach it was used several diamagnetic organic liquids which 
individual molecules, having a small number of interacted non-equivalent nuclear spins--qubits with $I = 1/2$ and being nearly 
independent one another. They act in parallel as an ensemble of 
{\it almost independent }quantum microcomputers.\par
Since in liquids the nuclear spins are very weakly coupled with 
the environment, the consideration may be restricted to nuclear spins 
of individual molecules ({\it reduced }quantum ensemble). For finite 
temperatures, the reduced ensemble is in the {\it mixed }state described by 
a density matrix.\par
{\it Initialization }of the nuclear spin states in this case means the 
separation of blocks from density matrix that must have properties 
similar to those of pure states (a pure-state diagonal matrix can 
have only one nonzero element). The states described by such matrix 
are called effective or {\it pseudo-pure }states. Several way of 
initialization have been reported in \cite{1,2,4,5}.\par
The access to individual qubits in liquid sample is replaced by 
simultaneous access to related qubits in all molecules of a bulk 
ensemble. Computers of this type are called {\it bulk-ensemble }quantum 
computers. They, in principle, can operate at {\it room temperature}.\par
For measurements of qubit states the standard NMR technique is 
used. A principle one-coil scheme of experiment is shown in Fig.1. 
The sample is placed in the constant external magnetic field $\mathbf{B}$ and
the alternating (say, linearly polarized) field $\mathbf{b}(t)$, produced by RF
voltage $V_{\omega }(t)$:
\begin{eqnarray}
\mathbf{B}(t) = \mathbf{B}\mathbf+ \mathbf{b}(t) = B \mathbf{k}\mathbf+ 2b \cos (\omega t + \varphi ) \mathbf{i}\mathbf,\label{1}
\end{eqnarray}
where $\mathbf{i}$ and $\mathbf{k}$ are the unit vectors along the axes x and z.
\par
\begin{center}
\epsfbox{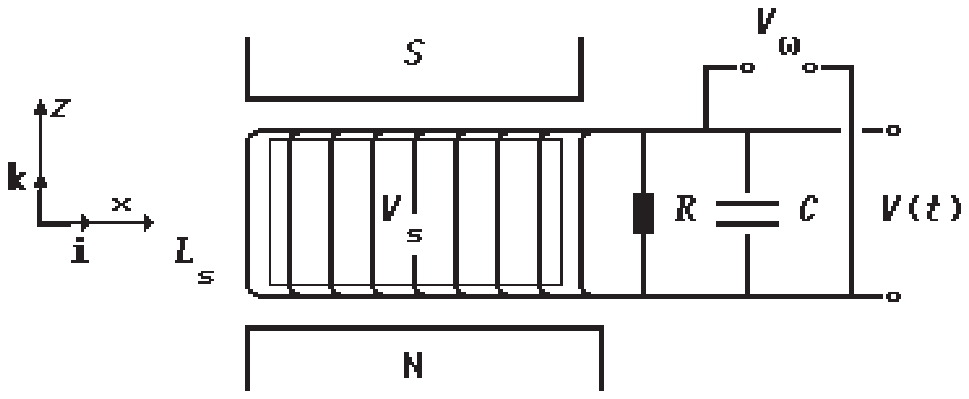}
\nobreak\par\nobreak
Fig. 1. The principle one-coil scheme of NMR measurement.\par
\end{center}
$\;$
\par
Let the sample be an ensemble of $N$ molecules--microcomputers
with $L$ qubits at temperature $T = 300\,\mathrm{K}$, in the external magnetic 
fields $B = 1-10\,\mathrm{T}$. The resonance nuclear spin frequency is 
$\omega _{\mathrm{A}}/2\pi \sim \gamma _{\mathrm{I}}B/2\pi < 150\,\mathrm{MHz}$, $\gamma _{\mathrm{I}}$ is gyromagnetic ratio of nuclear spin 
$(\gamma _{\mathrm{I}} \sim \gamma _{\mathrm{N}} = 95.8$ radMHz/T), $\hbar \omega _{\mathrm{A}}/kT < 10^{-5}$.\par
The output oscillating voltage $V(t))$ is
\begin{eqnarray}
V(t) = {\it QK d}\Phi (t)/dt = \mu _{0} {\it QK A dM}_{\mathrm{x}}(t)/dt ,\label{2}
\end{eqnarray}
where $\Phi (t) = \int_{A} \mu _{0}M_{\mathrm{x}}(t){\it dxdz}$ --- magnetic flux produced by resonant spins
in the coil, $L_{\mathrm{s}} = \mu _{0}(KA)^{2}/V$ --- solenoid inductance of the resonance 
counter $(\mu _{0} = 4\pi \cdot 10^{-1} \textrm{T}^{2}\,\mathrm{cm}^{3}/\mathrm{J})$, $V_{\mathrm{s}}$ --- volume of the solenoid, $K$ --- the
number and $A$ --- area of coil turns, for resonance $\omega = \omega _{\mathrm{A}} = (L_{\mathrm{s}}C)^{-1/2}$,
$Q = R/(\omega _{\mathrm{A}}L_{\mathrm{s}}) > 10^{2}$ --- the {\it quality factor} of resonance counter in 
parallel connected resistance $R$ (Fig.1).\par
The nuclear spin read-out magnetization $M_{\mathrm{x}}$ (the liquid sample is 
considered here as continuous medium) at optimum resonance conditions 
is defined by amplitude of RF field $b = 1/(\gamma _{\mathrm{I}} \sqrt{T_{\perp \mathrm{I}}T_{\parallel \mathrm{I}}})$. For 
transverse $T_{\perp \mathrm{I}}$ and longitudinal $T_{\parallel \mathrm{I}}$ nuclear spin relaxation times 
$\sim 10^{4}\,\mathrm{s}$ we have a value $b \sim 10^{-3}\,\mathrm{T}$ and \cite{6}
\begin{eqnarray}
M_{\mathrm{xmax}} = M_{\mathrm{z}0}\sqrt{T_{\perp \mathrm{I}}/T_{\parallel \mathrm{I}}}/2 \approx \gamma _{\mathrm{I}}\hbar /2\cdot (N/V_{\mathrm{s}})\cdot \varepsilon (L)/2 ,\label{3}
\end{eqnarray}
where $N$ --- number of resonant nuclear spins (one in a molecule) in 
volume $V_{\mathrm{s}}$. Parameter $\varepsilon (L)$ is the maximum probability of the full 
nuclear polarization in pseudo-pure state $P_{\mathrm{I}} = 1$ \cite{7}. It may be 
estimated by the difference of equilibrium population between the 
lowest and highest energy states. For nearly homonuclear $L$ spin 
system \cite{7} it is \footnote{We have used here the two different notation for resonance 
frequency $\omega _{\mathrm{A}}$ = $\gamma _{\mathrm{A}}B$ and $\omega _{\mathrm{I}}$ = $\gamma _{\mathrm{I}}B$. They determinate accordingly values 
$\varepsilon $ and $M_{\mathrm{z}0}$ in (\ref{2}). For homonuclear diamagnetic liquids
$\gamma _{\mathrm{A}}$ = $\gamma _{\mathrm{I}}$.}$^{)}$:\par
\begin{eqnarray}
\varepsilon (L) = \frac{\exp (L\hbar \omega _{\mathrm{A}}/2kT) - \exp (-L\hbar \omega _{\mathrm{A}}/2kT)}{(\exp (\hbar \omega _{\mathrm{A}}/2kT) + \exp (-\hbar \omega _{\mathrm{A}}/2kT))^{L}} = \frac{2\sinh(L\hbar \omega _{\mathrm{A}}/2kT)}{2^{L}\cosh ^{L}(\hbar \omega _{\mathrm{A}}/kT)}.\label{4}
\end{eqnarray}
The other estimations of $\varepsilon (L)$ are given in \cite{5}.\par
In the high temperature limit $\hbar \omega _{\mathrm{A}}/(kT) \ll 1$ we have 
$\varepsilon (L) = L2^{-L}\hbar \omega _{\mathrm{A}}/(kT)$, that is, the signal {\it exponentially drops }with the 
number of qubits, but {\it does not }for $\hbar \omega _{\mathrm{A}}/(kT) \gg 1$ when $\varepsilon (L) = 1$ (the 
pure nuclear quantum state).\par
The NMR signal intensity $S$ is defined by amplitude
\begin{eqnarray}
S = |V_{\mathrm{max}}| = (\mu _{0}/4) {\it Q KA }(N/V_{\mathrm{s}}) \gamma _{\mathrm{I}}\hbar \omega _{\mathrm{A}} \varepsilon (L) ,\label{5}
\end{eqnarray}
The product $KA$ can be expressed also as
\begin{eqnarray}
KA = (L_{\mathrm{s}}V_{\mathrm{s}}/\mu _{0})^{1/2} = (R V_{\mathrm{s}}/(\mu _{0}Q \omega _{\mathrm{A}}))^{1/2}.\label{6}
\end{eqnarray}
For the root-mean square noise voltage in the measurement circuit we become
\begin{eqnarray}
V_{\mathrm{N}} = \sqrt{4{\it kT R }\Delta \nu } ,\label{7}
\end{eqnarray}
where $\Delta \nu \sim 1\,\mathrm{Hz}$ --- the amplifier bandwidth.\par
So that for {\it signal to noise ratio} we obtain\par
\begin{eqnarray}
(\mathrm{S}/\mathrm{N}) \equiv |V_{\mathrm{max}}|/V_{\mathrm{N}} \cong \frac{1}{8} \sqrt{ \frac{\mu _{0}\hbar Q \hbar \omega _{\mathrm{A}}}{V_{\mathrm{s}} \Delta \nu kT}} \gamma _{\mathrm{I}}N \varepsilon (L) \sim \nonumber
\end{eqnarray}\begin{eqnarray}
\sim \sqrt{(Q/V_{\mathrm{s}})\cdot (\hbar \omega _{\mathrm{A}}/kT)} N \varepsilon (L)\cdot 10^{-9}, (\mathrm{here}\;V_{\mathrm{s}}\;\mathrm{in}\;\mathrm{cm}^{3}).\label{8}
\end{eqnarray}
For example, for two qubits molecules $(L = 2)$, using, 
$\varepsilon (L) = \hbar \omega _{\mathrm{A}}/(kT) \sim 10^{-5}$, we can make an estimate\begin{eqnarray}
(\mathrm{S}/\mathrm{N}) \sim (Q/V_{\mathrm{s}})^{1/2} N 10^{-17}.\label{9}
\end{eqnarray}
Thus, the two qubits liquid ensemble at room temperature for 
$V_{\mathrm{s}} \sim 1\,\mathrm{cm}^{3}$ and $Q \sim 10^{3}$ is bounded by $N > 10^{16} $molecules.\par
In the case of {\it paramagnetic liquids }one would expect that the 
number of molecules may be more increased with the use of a dynamic 
polarization (say, Overhauser effect). Assuming for electron and 
nuclear gyromagnetic ratio $\gamma _{\mathrm{e}}/\gamma _{\mathrm{I} }\sim 10^{3}$ we obtain that in the 
probability $\varepsilon (L)$ for a $L - $qubits single state the value 
$\hbar \omega _{\mathrm{A}}/(kT) \ll 10^{3}$ should be replaced by 10$^{3}\hbar \omega _{\mathrm{A}}/(kT)$. Therefore for the 
same value $\varepsilon (L)$ the allowed number of qubits $L$ approximately will be 
estimated from
\begin{eqnarray}
L2^{-L} > 10^{-3} ,\label{10}
\end{eqnarray}
whence it follows that $L < 12$ qubits.\par
An additional increasing of read-out NMR signal in paramagnetic 
liquids may be obtained using the ENDOR technique. It is generally 
believed that for the liquid bull-ensemble quantum computers a 
{\it limiting value} is $L < 20-30$ \cite{7}.\par
However, for realization of {\it large-scale NMR quantum computers }a 
necessary number of qubits in quantum register must be $L > 10^{3}$, that 
may be attained only {\it at low temperatures in solid-state }structures.\par
The radically new and still unimplemented design of solid-state 
quantum computer was proposed in \cite{8,9}. It involves the formation of 
an multiple-spin system with access to {\it individual }nuclear spins--qubits.
It was suggested to use an semiconductor MOS structure on a 
$^{28}\mathrm{Si}$ spinless substrate, into a near-surface layer of which $^{31}\mathrm{P}$ 
stable phosphorus isotopes, acting as donors, are implanted in form 
of regular chain. These donors have a nuclear spin $I = 1/2$ and 
substitute for silicon atoms at the lattice sites, producing shallow 
impurity states with a large effective Bohr radius. The number of 
donors or the qubit number $L$ in a such quasi-one-dimensional
{\it artificial ''molecule'' }may be arbitrary large. It is suggested an 
electrical control and measurement of qubits states through the use 
of special gate structures.\par
The experimental implementation of Kane scheme is undertaken now 
in Australian Centre for Quantum Computer Technology \cite{10}.\par
There are four essential difficulties in implementing of quantum 
computer that was suggested by B. Kane in \cite{8,9}:\par
1) First of all, signal from the spin of an individual atom is 
very small and it is required of {\it high sensitive single-electron 
measurements}.\par
2) For initialization nuclear spin states it is required to use 
of {\it very low nuclear spin temperature} $(\sim \,\mathrm{mK})$.\par
3) In is required of regular donors and gates arrangement with 
high precision in {\it nanometer scale}.\par
4) It is necessary to suppress the {\it decoherence }of quantum states 
defined by {\it thermal fluctuations }of gate voltage.\par
As an alternative, the variant of {\it an ensemble silicon-based 
quantum computer} was proposed by us \cite{11,12}. One would expect that 
with the ensemble approach, where many independent ''molecules'' of 
Kane's type work simultaneously, the measurements would be greatly 
simplified. Here we will give a some further development of the 
scheme considered in \cite{11,12}.\par
\par
\section{The silicon structure with regular system of strip gates}
\par
In this case, unlike the structure suggested in \cite{8}, gates $\mathbf{A}$ and 
$\mathbf{J}$ form a chain of narrow $(l_{\mathrm{A}} \sim 10\,\mathrm{nm})$ and long strips along which 
donor atoms at $l_{\mathrm{y}}$ distant from each other are placed (Fig.2). Thus, 
they form a {\it regular }structure of the planar silicon topology type.\par
The separation between neighbourig donor atoms in Si, as in
Kane's scheme, must be $l_{\mathrm{x}} \leq  20\,\mathrm{nm})$. In this case the interqubit
interaction is controlled by gates $\mathbf{J}$. The depth of donor occurring
$d \sim 20\,\mathrm{nm}$. For $l_{\mathrm{y}} \gg l_{\mathrm{x}}$ the exchange spin interaction between
electrons of donor atoms along the strip gates ($y-$axis) is assumed
negligibly small. Then, such a system breaks down into an ensemble of
near-independent chains--Kane's artificial ''molecule, whose
electronic spins at temperature $T \leq 0,1\,\mathrm{K}$ are initially fully aligned
with the field of several Tesla $(\gamma _{\mathrm{e}}\hbar B/kT \gg 1)$. The linear qubit
density in artificial ''molecules'' is $\sim 50$ qubits on micrometer.\par
For {\it the initializing }of all nuclear spins-qubit quantum states
(fully polarized nuclear spins) there is a need to attain, for the
time being, spin temperature $T \leq 10^{-3}\,\mathrm{K}$. An output signal in this
system, as in liquids, will be proportional to the number of the
''molecules'' or donor atoms $N$ (components of our ensemble) in the
chain along axis $y$.\par
\par
\begin{center}
\epsfbox{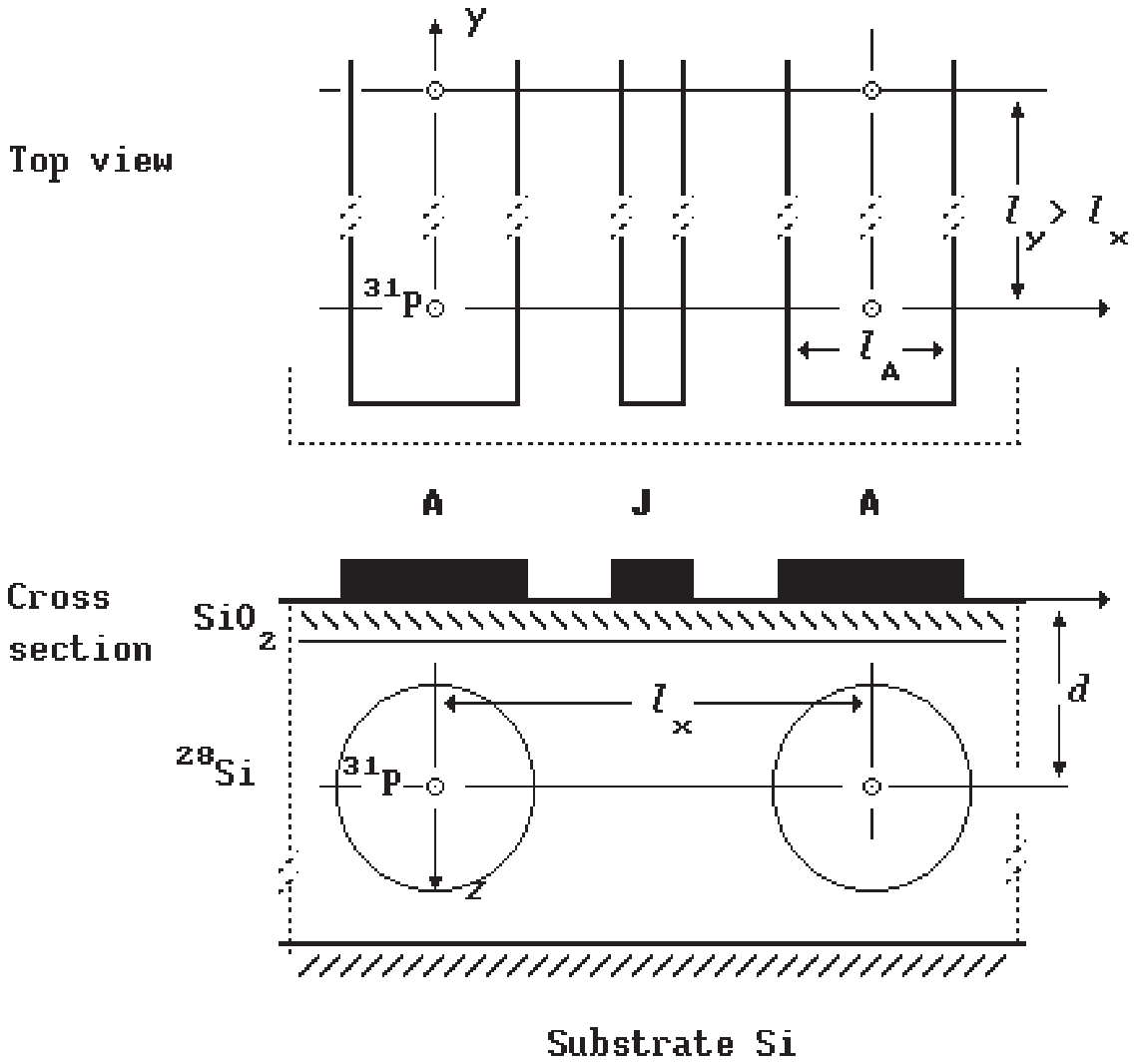}
\nobreak\par\nobreak
Fig. 2. The silicon structure $(^{28}\mathrm{Si})$ with donors atoms $^{31}\mathrm{P}$ and
strip gates (Al).\par
\end{center}
$\;$
\par
However for the realization of considered structure with donor
chains and parallel strip gates, as well as of the Kane's scheme, {\it the
nanotechnology }with resolving of order $\sim $ 1 םל is also needed. We will
not discuss here the technology problems in realizing of the
suggested silicon-based structure (see \cite{10}).\par
\par
\section{The states of insulated donor atoms in magnetic fields}
\par
The electron-nuclear spin Hamiltonian for a donor atom $^{31}\mathrm{P}$ has 
the form\begin{eqnarray}
\hat{H} = \gamma _{\mathrm{e}}\hbar \mathbf{B}\hat{\mathbf{S}} - \gamma _{\mathrm{I}}\hbar \mathbf{B}\hat{\mathbf{I}} + A \hat\mathbf{I}\hat\mathbf{S} ,\label{11}
\end{eqnarray}
four energy levels of which are given by the well-known Breit-Rabi's 
formula. For $I = 1/2$, $S = 1/2$ (the $z-$axis is parallel to $\mathbf{B})$ this 
formula is written as\begin{eqnarray}
E(F,m_{\mathrm{F}}) = - \frac{A}{4} - \gamma _{\mathrm{I}}\hbar Bm_{\mathrm{F}} - (-1)^{\mathrm{F}} \frac{A}{2} \sqrt{1 + 2m_{\mathrm{F}}X + X^{2}} ,\label{12}
\end{eqnarray}
where constant of hyperfine interaction $A/(2\pi \hbar ) = 116\,\mathrm{MHz} \cite{13}$, 
$X = (\gamma _{\mathrm{e}} + \gamma _{\mathrm{I}})\hbar B/A \approx \gamma _{\mathrm{e}}\hbar B/A$, $F = I \pm 1/2 = 1, 0$, and 
$m_{\mathrm{F}} = M + m = \pm 1, 0$, if $F = 1$ or $m_{\mathrm{F}} = 0$, if $F = 0$ (Here $M = \pm 1/2$ 
and $m = \pm 1/2$ are $z-$projections of electronic and nuclear spins 
accordingly). The energy levels are shown in Fig. 3. For the energy 
of the ground spin state, $F = 0$ and $m_{\mathrm{F}} = 0$, hence, we obtain\begin{eqnarray}
E(0,0) = - A/4 - (A/2) \sqrt{1 + X^{2}}.\label{13}
\end{eqnarray}
For the next, excited energy state, $F = 1$, $m_{\mathrm{F}} = -1$ we have\begin{eqnarray}
E(1,-1) = A/4 - (\gamma _{\mathrm{e}} - \gamma _{\mathrm{I}}) \hbar B/2.\label{14}
\end{eqnarray}
Thus, the energy difference between the two states of the nuclear 
spin that interacts with an electron, whose state remains unchanged, 
is described in simple terms $(\gamma _{\mathrm{e}} \gg \gamma _{\mathrm{I}}$ for $X \approx \gamma _{\mathrm{e}}\hbar B/A \gg 1)$:\begin{eqnarray}
\hbar \omega _{\mathrm{A}}^{+} = E(1,-1) - E(0,0) = A/2 + (\gamma _{\mathrm{I}} - \gamma _{\mathrm{e}})\hbar B/2 + \frac{A}{2} \sqrt{1 + X^{2}} \approx \nonumber
\end{eqnarray}\begin{eqnarray}
\approx \gamma _{\mathrm{I}}\hbar B + \frac{A}{2} - \frac{A^{2}}{4\gamma _{\mathrm{e}}\hbar B} ,\nonumber
\end{eqnarray}\begin{eqnarray}
\hbar \omega _{\mathrm{A}}^{-} = E(1,1) - E(1,0) \approx - \gamma _{\mathrm{I}}\hbar B + \frac{A}{2} + \frac{A^{2}}{4\gamma _{\mathrm{e}}\hbar B}.\label{15}
\end{eqnarray}
For $^{31}\mathrm{P}$ donor atoms $\gamma _{\mathrm{e}}/\gamma _{\mathrm{I}} = 1.62\cdot 10^{3}$, $\gamma _{\mathrm{e}} = 176.08$ radGHz/T, 
$\gamma _{\mathrm{I}} = 1.13\gamma _{\mathrm{N}} = 108$ radMHz/T. In magnetic field $B = 1\,\mathrm{T}$: 
$\omega _{\mathrm{A}}^{+}/2\pi = 75\,\mathrm{MHz}$, $\omega _{\mathrm{A}}^{-}/2\pi = 41\,\mathrm{MHz}$.\par
\par
\begin{center}
\epsfbox{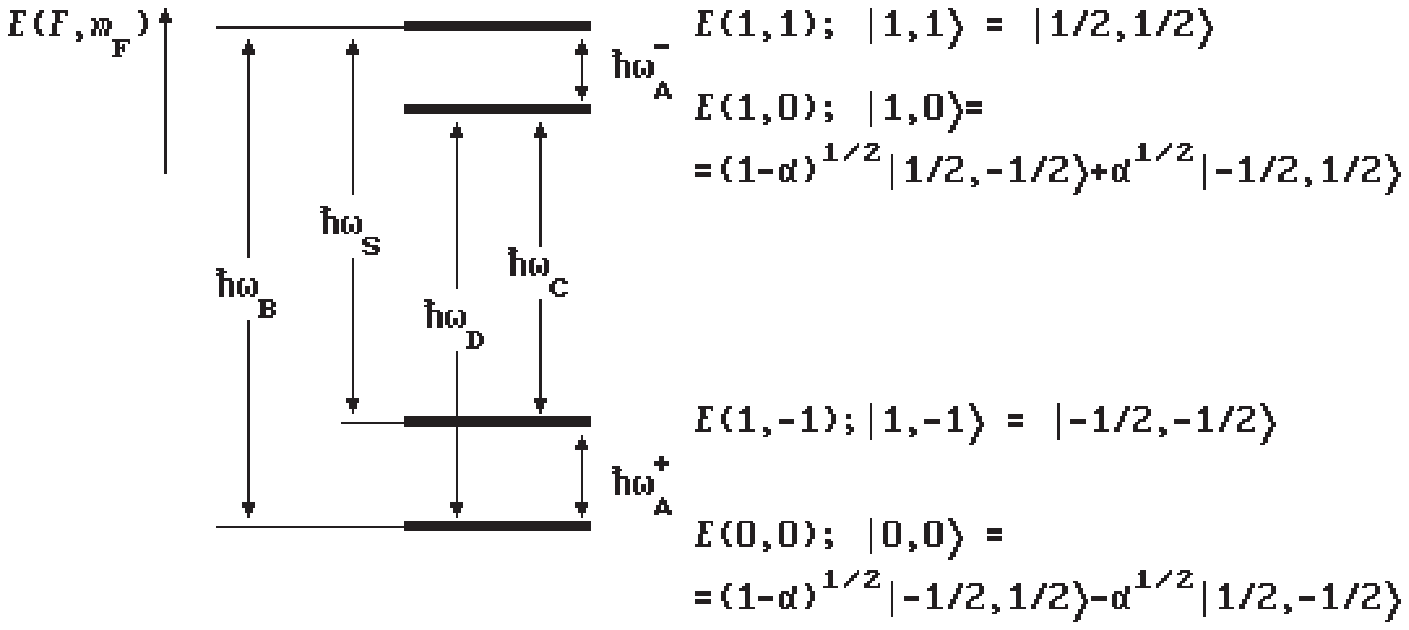}
\nobreak\par\nobreak
Fig. 3. Energy levels of an individual donor atom in magnetic
field. The frequencies $\omega _{\mathrm{S}} = \omega _{\mathrm{B}} - \omega _{\mathrm{A}}^{+}$, $\omega _{\mathrm{B}}$, $\omega _{\mathrm{C}}$, $\omega _{\mathrm{D}} = \omega _{\mathrm{B}} - \omega _{\mathrm{A}}^{-}$ are in
microwave, $\omega _{\mathrm{A}}^{\pm }$ -- in the RF range of frequencies. The transitions with
frequencies $\omega _{\mathrm{B}}$, $\omega _{\mathrm{D}}$, $\omega _{\mathrm{C}}$, $\omega _{\mathrm{A}}^{\pm }$ are {\it allowed}, but transitions with
frequencies $\omega _{\mathrm{S}}$ in the first approximation are {\it forbidden}.\par
\end{center}
$\;$
\par
The state $|1,-1\rangle $ in $M, m$ basis is equivalent to state 
$|M = -1/2, m = -1/2\rangle $ and state $|0, 0\rangle $ in this basis is the following 
singlet superposition of states\begin{eqnarray}
|0,0\rangle = (1 - \alpha )^{1/2} |-1/2, 1/2\rangle - \alpha ^{1/2} |1/2, -1/2\rangle ,\nonumber
\end{eqnarray}\begin{eqnarray}
\alpha = \frac{1}{2} \left(1 - \frac{X}{\sqrt{1 + X^{2}} }\right)\label{16}
\end{eqnarray}
The diagonal matrix elements of nuclear magnetization $M_{\mathrm{z}0}$ for 
two lower energy states will be determinated by\begin{eqnarray}
\langle 0,0|M_{\mathrm{z}0}|0,0\rangle = \langle 0,0|I_{\mathrm{z}}|0,0\rangle \gamma _{\mathrm{I}}\hbar = \frac{X}{\sqrt{1 + X^{2}} } \gamma _{\mathrm{I}}\hbar /2 ,\nonumber
\end{eqnarray}\begin{eqnarray}
\langle 1,-1|M_{\mathrm{z}0}|1,-1\rangle = \langle 1,-1|I_{\mathrm{z}}|1,-1\rangle \gamma _{\mathrm{I}}\hbar = - \gamma _{\mathrm{I}}\hbar /2.\label{17}
\end{eqnarray}
The populations of the $L - $qubit lowest and highest energy states for 
$M = - 1/2$, which give, as noted above, the maximum probability of the 
full nuclear polarisation in pseudo-pure state, are:\begin{eqnarray}
p(1,-1) = \frac{\exp (-L\hbar \omega _{\mathrm{A}}^{+}/2kT)}{(\exp (\hbar \omega _{\mathrm{A}}^{+}/2kT) + \exp (-\hbar \omega _{\mathrm{A}}^{+}/2kT))^{L}} ,\nonumber
\end{eqnarray}\begin{eqnarray}
p(0,0) = \frac{\exp (L\hbar \omega _{\mathrm{A}}^{+}/2kT)}{(\exp (\hbar \omega _{\mathrm{A}}^{+}/2kT) + \exp (-\hbar \omega _{\mathrm{A}}^{+}/2kT))^{L}}.\label{18}
\end{eqnarray}
The maximum nuclear magnetization $M_{\mathrm{z}0}$ (the populations of states 
(1,1) and (1.0) are negligible for $\omega _{\mathrm{S}}, \omega _{\mathrm{B}}, \omega _{\mathrm{C}} \gg \omega _{\mathrm{A}}^{\pm })$ is
\begin{eqnarray}
M_{\mathrm{z}0} = \gamma _{\mathrm{I}}\hbar /2\cdot (N/V_{\mathrm{c}}) \left\{{ \frac{X}{\sqrt{1 + X^{2}} } \frac{\exp (L\hbar \omega _{\mathrm{A}}^{+}/2kT)}{(\exp (\hbar \omega _{\mathrm{A}}^{+}/2kT) + \exp (-\hbar \omega _{\mathrm{A}}^{+}/2kT))^{L}} -}\right.
  \nonumber
\end{eqnarray}
\begin{eqnarray}
\left.{ - \frac{\exp (-L\hbar \omega _{\mathrm{A}}^{+}/2kT)}{(\exp (\hbar \omega _{\mathrm{A}}^{+}/2kT) + \exp (-\hbar \omega _{\mathrm{A}}^{+}/2kT))^{L}} }\right\}.\label{19}
\end{eqnarray}
For $L\hbar \omega _{\mathrm{A}}^{+}/2kT \ll 1$ and $X \gg 1$ we obtain (compare with (\ref{3}))\begin{eqnarray}
M_{\mathrm{z}0} \approx \gamma _{\mathrm{I}}\hbar /2\cdot (N/V_{\mathrm{c}})\cdot 2^{-L}L(\hbar \omega _{\mathrm{A}}^{+}/kT).\label{20}
\end{eqnarray}
But for very low temperatures $(\hbar \omega _{\mathrm{A}}^{+}/2kT \gg 1)$ we have the {\it full nuclear 
polarization}: $M_{\mathrm{z}0} \approx \gamma _{\mathrm{I}}\hbar /2\cdot (N/V_{\mathrm{c}})$ and $\varepsilon (L) = 1$.\par
\par
\section{The gain effect for NMR signal}
\par
Transitions between two lower states are induced by a RF 
magnetic field, applied at a frequency resonant $\omega _{\mathrm{A}}$. The Rabi 
resonance frequency $\Omega $, which is defined by matrix elements of spin 
interaction Hamiltonian with the external RF field $\mathbf{b}(t)$\begin{eqnarray}
\hat{H}_{\mathrm{rf}} = \left( \gamma _{\mathrm{e}}\hat{S}_{\mathrm{x}} - \gamma _{\mathrm{I}}\hat{I}_{\mathrm{x}}\right)\hbar \cdot b_{\mathrm{x}}(t) ,\label{21}
\end{eqnarray}can be found from\begin{eqnarray}
\Omega = \gamma _{\mathrm{I}}b_{\mathrm{eff}}(X) = 2 |\langle 0,0|\hat{H}_{\mathrm{rf}}|1,-1\rangle |/\hbar .\label{22}
\end{eqnarray}
For the amplitude of effective RF field, acting on nuclear spin, 
$b_{\mathrm{eff}}(X)$ we obtain
\begin{eqnarray}
b_{\mathrm{eff}}(X) = b \left( \alpha ^{1/2} \gamma_{\mathrm{e}}/\gamma_{\mathrm{I}} + (1 - \alpha )^{1/2} \right) ,\label{23}
\end{eqnarray}
where $b$ --- the amplitude of circularly polarized field component.\par
The Rabi frequency has the maximum value for $X = 0$ and rapidly 
monotonic reduce to value for insulated nuclear spin $(A \Rightarrow 0)$, 
$\gamma _{\mathrm{I}}b_{\mathrm{eff}}(X) \gg 1) = \gamma _{\mathrm{I}}b$. From a rate of quantum operation standpoint 
it is desirable to operate in the relatively weak fields, at which 
$\gamma _{\mathrm{e}}/\gamma _{\mathrm{I}} \gg X \approx \gamma _{\mathrm{e}}\hbar B/A \gg 1$ or $3.5 T > B \gg 3.9\cdot 10^{-3}\,\mathrm{T}$.\par
In this case from (\ref{23}) we will obtain
\begin{eqnarray}
b_{\mathrm{eff}} = (1 + \eta )b \gg b,\label{24}
\end{eqnarray}where $\eta = A/(2\gamma _{\mathrm{I}}\hbar B) \gg 1$ is the {\it gain factor}. On these conditions RF 
field operate through the transverse component of electronic 
polarization. For magnetic fields $B = 1\,\mathrm{T}$ we have the value 
$b_{\mathrm{eff}} = 4.4\cdot b$, and for $B = 0.01\,\mathrm{T}$ -- value $b_{\mathrm{eff}} = 338\cdot b$.\par
The gain effect involves an increase of NMR signal and Rabi 
frequency. In the pulse technique it make possible {\it to decrease the 
length of pulse}. This effect was first indicated by Valiev in \cite{14}. 
The computer operations, owing to this effect, can be done at lower 
RF fields. At last, it permits to {\it reduce the RF field influence} on 
the operation of neighbour semiconductor devices.\par
To describe the nuclear dynamics of the studied two low-lying 
levels system we can write the followed Bloch-type equation:\par
\begin{eqnarray}
\frac{d\mathbf{M}}{dt} = \gamma _{\mathrm{I}} [\mathbf{M}\mathbf\times \mathbf{B}_{\mathrm{eff}}] - \frac{M_{\mathrm{x}}\mathbf{i}\mathbf+ M_{\mathrm{y}}\mathbf{j}}{T_{\perp \mathrm{I}}} - \frac{(M_{\mathrm{z}} - M_{\mathrm{z}0})\mathbf{k}}{T_{\parallel \mathrm{I}}} ,\label{25}
\end{eqnarray}
where $\mathbf{i}$, $\mathbf{j}$, $\mathbf{k}$ are orthogonal unit vectors (Fig.1),
\begin{eqnarray}
\mathbf{B}_{\mathrm{eff}} = \left(\omega _{\mathrm{A}}/\gamma _{\mathrm{I}} \right) \mathbf{k}\mathbf+ 2b_{\mathrm{eff}}\cos (\omega t) \mathbf{i}\mathbf,\label{26}
\end{eqnarray}
from where it follows that the value of maximum nuclear read-out
magnetization in NMR signal, which there have been here for
$b_{\mathrm{eff}}(X) = 1/(\gamma _{\mathrm{I}}\sqrt{T_{\perp \mathrm{I}}T_{\parallel \mathrm{I}}})$, again have value $M_{\mathrm{x}} = M_{\mathrm{z}0}/2$. That is, the
read-out {\it NMR signal can not be increased }through the gain effect over 
its maximum value $M_{\mathrm{z}0}/2$, but it {\it can be achieved at more lower }RF 
fields than in the absent of gain effect.\par
\par
\section{The signal to noise ratio for an ensemble silicon quantum computer}
\par
For the realization of an ensemble silicon quantum register we 
propose a variant of planar scheme, that contains $n\cdot p$ in parallel 
acting blocks, each have $N_{0}$ in parallel connected $L-$qubit Kane's 
linear ''molecules''. This scheme is schematically depicted at Fig. 4.\par
Let the sample be the silicon $(^{28}\mathrm{Si})$ plate of thick $\delta \sim 0.1\,\mathrm{cm}$. 
For the full number of computers-''molecules'' in ensemble $N = p\cdot N_{0}\cdot n$, 
the volume of sample and also of solenoid is $V_{\mathrm{s}} \approx \delta \cdot l_{\mathrm{x}}\cdot l_{\mathrm{y}}\cdot L\cdot N$ (the 
filling factor is assumed to be approximately one).\par
\par
\begin{center}
\epsfbox{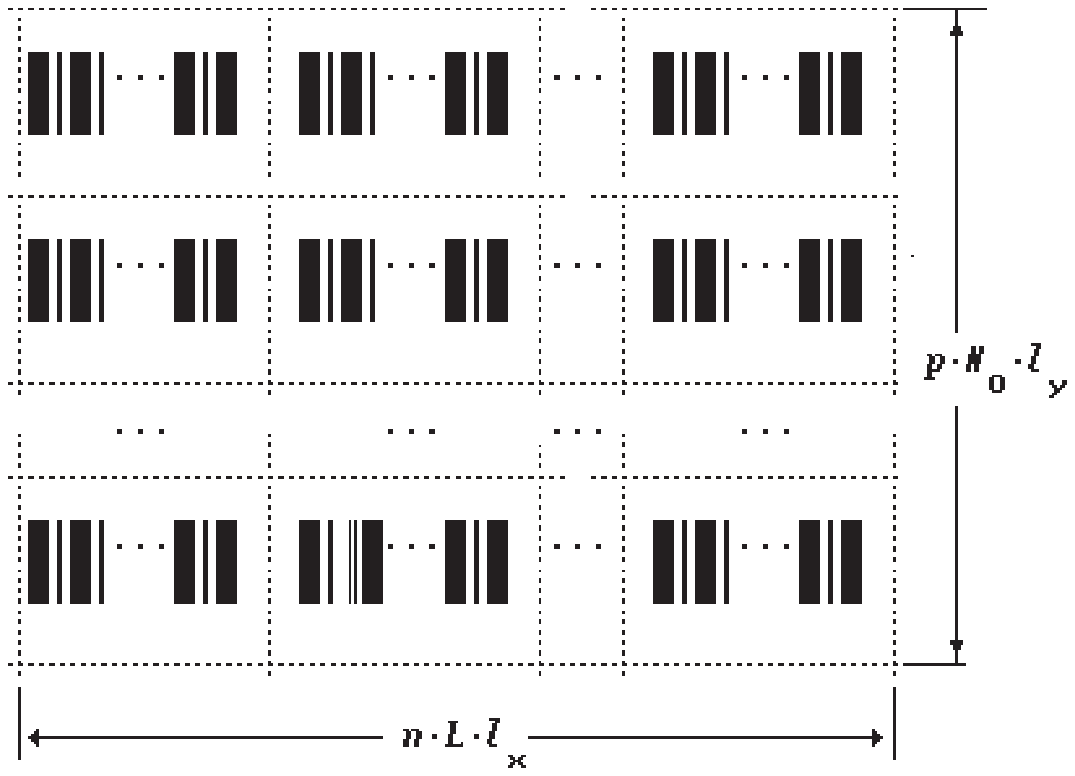}
\nobreak\par\nobreak
Fig. 4. The scheme of the planar silicon topology with $p\cdot n$ in
parallel connected blocks of the ensemble $L - $qubit quantum computers
(the connections are not shown here). The broad and narrow lines
denote the $\mathbf{A}$ and $\mathbf{J}$ gates.\par
\end{center}
$\;$
\par
The read-out signal from the such {\it ensemble in parallel acting 
}chains, as distinct from liquid prototype, for full nuclear 
polarization or, what is the same, for spin temperatures $T_{\mathrm{I}}$ $\leq $ 10$^{-3}\,\mathrm{K}$ 
has no small factor in intensity of the NMR signal of type 
$\varepsilon (L) = 2^{-L}L\cdot \hbar \omega _{\mathrm{A}}/(kT)$. The NMR signal from our sample within of a non-essential factor is the same as from macroscopic sample (see 
Appendix). Therefore, with the expression (\ref{8}), $\hbar \omega _{\mathrm{A}}/(kT) \sim 10^{-2}$ 
$(T \sim 0.1\,\mathrm{K})$ and $\varepsilon = 1$ we will obtain as an estimation\begin{eqnarray}
(\mathrm{S}/\mathrm{N}) \approx \sqrt{Q\hbar \omega _{\mathrm{A}}/(kTV_{\mathrm{s}})}\cdot N\cdot 10^{-9} \approx \sqrt{QN/(\delta l_{\mathrm{x}}l_{\mathrm{y}}L)}\cdot 10^{-10}.\label{27}
\end{eqnarray}
It is believed that for superconducting Al gates it is possible 
$Q \sim 10^{6}$. The effective volume of one ''molecule'' for $l_{\mathrm{x}} = 20\,\mathrm{nm}$, 
$l_{\mathrm{y}} = 50\,\mathrm{nm}$, $L = 10^{3}$, $V_{\mathrm{s}} = \delta l_{\mathrm{x}}l_{\mathrm{y}}L = 10^{-9}\,\mathrm{cm}^{3}$ we receive that the read-out signal in our scheme may be {\it available for standard NMR technique 
}if the number of ''molecules'' in ensemble will be $N \geq 10^{5}$. High-sensitive devices for individual measurements are not needed.\par
To estimate the values $n$, $p$ let us consider the square sample 
with $50 N_{0}p = 20\cdot 10^{3} n$ and $N_{0} = 100$. As a result, we receive $n \approx 16$ 
and $p \approx 63$. The area of the structure is $\sim 315 \times 315 \,\mu\mathrm{m}^{2}$. This size 
is sufficiently small for sample to be housed in the split between 
the magnet poles of a NMR spectrometer.\par
For implementation of two-qubit logic operation it is required 
the controlled by gates $\mathbf{J}$ interqubit indirect interaction with 
characteristic frequency $\nu _{\mathrm{J}} \sim 100\,\mathrm{kHz} \ll \omega _{\mathrm{A}}/2\pi \sim 100\,\mathrm{MHz}$. To bring 
about fault-tolerant quantum computations the relative error for 
single logic operation must be no more then $\sim $ 10$^{-5}$ \cite{15}. Hence it 
follows that a {\it resolution }bound of the NMR spectrometer must be of 
order $\sim 100\,\mathrm{kHz}\cdot 10^{-5} \sim 1\,\mathrm{Hz}$, that is consistent with the usual 
requirements. Notice that such high precision is needed only for 
performing the logic quantum operation, but it is not needed for 
read-out measurements.\par
The read-out signal may be some more increased by means of an 
electron-nuclear double resonance (ENDOR) methods \cite{16} of observing 
the electron resonance at transition with frequencies $\omega _{\mathrm{B}}$ and $\omega _{\mathrm{C}}$ 
(Fig. 3).\par
Consequently, by use of standard NMR and additional of ENDOR 
techniques {\it the first main difficulty }of Kane's scheme can be 
overcame.\par
\par
\section{The longitudinal relaxation times and long-lived nonequilibrium quantum states}
\par
The electron and nuclear {\it longitudinal relaxation times }for the 
{\it allowed }transitions in four energy levels system of phosphorus doped 
silicon have been extensively investigated experimentally in \cite{16}, 
see also \cite{17}. For the allowed transitions with frequency $\omega _{\mathrm{B}}$ and $\omega _{\mathrm{C}}$ 
(Fig.3) electron longitudinal relaxation times $T_{\parallel \mathrm{B}} \approx T_{\parallel \mathrm{C}}$ at low 
temperatures was found to be exceedingly long. They are of order {\it one 
hour }at $T = 1.25\,\mathrm{K}$, $B \sim 0,3\,\mathrm{T}$, are independent of phosphorus 
concentration below $C \sim 10^{16}\,\mathrm{cm}^{-3}$ and are approximately inversely 
proportional to lattice temperature $T$. But they very fast drops to 
$\sim 1\,\mathrm{s}$ already at $C \sim 10^{17}\,\mathrm{cm}^{-3}$. For the structure on Fig.2 $d = 20\,\mathrm{nm}$, 
$l_{\mathrm{x}} = 20\,\mathrm{nm}$, $l_{\mathrm{y}} = 50\,\mathrm{nm}$ we obtain the average distance between atoms 
$^{31}\mathrm{P}$ in silicon $l_{\mathrm{av}} = (20\cdot 20\cdot 50)^{1/3} = 27\,\mathrm{nm}$ and their associated 
concentration $C = (l_{\mathrm{av}})^{-3} = 5.10^{16}\,\mathrm{cm}$. For this value of $C$ according 
to \cite{16} we must have $T_{\parallel \mathrm{B}} \sim 100\,\mathrm{s}$.\par
The further increasing the electron relaxation times may be 
achieved by decreasing of lattice temperature. It is possible to 
extrapolate electron relaxation time to $T \sim 0.1\,\mathrm{K}$ resulting in a 
value $T_{\parallel \mathrm{B}} \sim 1000\,\mathrm{s}$.\par
The relaxation time for transition with frequency $\omega _{\mathrm{D}}$, which 
involves a simultaneous electron-nuclear spin flip-flop, at 
$T = 1.25\,\mathrm{K}$, $C \sim 10^{16}\,\mathrm{cm}^{-3}$ and $B \sim 0,3\,\mathrm{T}$ was 
$T_{\parallel \mathrm{D}} \sim 30$ hours $\gg T_{\parallel \mathrm{B}}, T_{\parallel \mathrm{C}}$.\par
The nuclear longitudinal relaxation time $T_{\parallel \mathrm{I}}$ exceed 10 hours 
\cite{15}. Notice, that {\it for full electronic polarization}, occurring at 
$T \leq 0,1\,\mathrm{K}$, this time {\it does not depend on the interaction with 
neighbour electron spins }\cite{17}.\par
The dipole-dipole interaction of separated qubits in our chains 
with polarized electron spins of neighbours phosphorus atoms is 
divided into {\it secular }and {\it nonsecular }parts. The main part of it is the 
secular part. It lead up to constant shift of resonance nuclear 
frequencies of order $\mu _{0}\gamma _{\mathrm{N}}\gamma _{\mathrm{B}}\hbar /(4\pi l^{3}_{\mathrm{x}}) \sim 10\,\mathrm{Hz}$, where
$\gamma _{\mathrm{B}} = 2\mu _{\mathrm{B}}/\hbar = 176\,\mathrm{radGHz}/\mathrm{T}$, $l_{\mathrm{x}} \sim 20\,\mathrm{nm}$.\par
To suppress the influence of the unwanted secular part of 
dipole-dipole spin interaction in a chain, may be applied the 
external magnetic field, $\mathbf{B}$, at an ''magic'' angle, 
$\theta = $arccos(1/$\sqrt{3}) = 54.74^{0}$ to direction of the chain.\par
The {\it extremely long relaxation times }of the electron and nuclear 
spins imply that the required initializing of nuclear quantum states 
(full nuclear {\it nonequilibrium polarizations}) can be attained by deep 
cooling of short duration {\it only nuclear spin system }to $T \leq 1\,\mathrm{mK}$ 
without deep cooling of lattice. There is the possibility to reach it 
at {\it indirect cooling }of nuclear spin system by means of {\it dynamic 
nuclear spin polarization techniques }\cite{17}.\par
\par
\section{The cooling of nuclear spin system and initialization nuclear 
states by means of dynamic polarization}
\par
One such method of dynamic nuclear spin polarization for donor 
atoms is based on the saturation by {\it the microwave pumping of the 
forbidden transition }(frequency $\omega _{\mathrm{S}}$ in Fig.3), that is designated as 
the {\it solid state effect }\cite{6,17}.\par
Let us consider this effect. The polarizations of electrons 
$P_{\mathrm{S}} = 2\langle S_{\mathrm{z}}\rangle $ and of nuclei $P_{\mathrm{I}} = 2\langle I_{\mathrm{z}}\rangle $ can be expressed as\par
\begin{eqnarray}
P_{\mathrm{S}} = p(1,1) + p(1,0) - p(1,-1) - p(0,0) ,\label{28}
\end{eqnarray}\begin{eqnarray}
P_{\mathrm{I}} = p(1,1) + p(0,0) - p(1,0) - p(1,-1) ,\nonumber
\end{eqnarray}
where $p(F,M_{\mathrm{F}})$ are the populations of states $F,M_{\mathrm{F}}$ (Fig.3). They fill 
also the requirement\begin{eqnarray}
p(1,1) + p(1,0) + p(1,-1) + p(0,0) = 1.\label{29}
\end{eqnarray}
Let us write the rate equations for the populations:\par
\begin{eqnarray}
dp(0,0)/dt = (p(1,1) - p(0,0)r_{\mathrm{B}})/T_{\parallel \mathrm{B}}+ (p(1,0) - p(0,0)r_{\mathrm{D}})/T_{\parallel \mathrm{D}}+\nonumber
\end{eqnarray}\begin{eqnarray}
+ (p(1,-1) - p(0,0)r_{\mathrm{A}}^{+})/T_{\parallel \mathrm{A}} ,\nonumber
\end{eqnarray}\begin{eqnarray}
dp(1,-1)/dt = (p(1,0) - p(1,-1)r_{\mathrm{C}})/T_{\parallel \mathrm{C}} + (p(1,1) - p(1,-1))\cdot W_{\mathrm{e}}+\nonumber
\end{eqnarray}\begin{eqnarray}
+ (p(0,0)r_{\mathrm{A}}^{+} - p(1,-1))/T_{\parallel \mathrm{A}} ,\nonumber
\end{eqnarray}\begin{eqnarray}
dp(1,0)/dt = (p(1,-1)r_{\mathrm{C}}- p(1,0))/T_{\parallel \mathrm{C}} + p(0,0)r_{\mathrm{D}}-p (1,0))/T_{\parallel \mathrm{D}} +\nonumber
\end{eqnarray}\begin{eqnarray}
+ (p(1,1) - p(1,0)r_{\mathrm{A}}^{-})/T_{\parallel \mathrm{A}} \nonumber
\end{eqnarray}\begin{eqnarray}
dp(1,1)/dt = (p(0,0)r_{\mathrm{B}} - p(1,1))/T_{\parallel \mathrm{B}} + (p(1,-1) - p(1,1))\cdot W_{\mathrm{e}} +\nonumber
\end{eqnarray}\begin{eqnarray}
+ (p(1,0)r_{\mathrm{A}}^{-} - p(1,1))/T_{\parallel \mathrm{A}}\label{30}
\end{eqnarray}
where parameters $r_{\mathrm{B},\mathrm{C},\mathrm{D},\mathrm{A}} =\,\exp (-\hbar \omega _{\mathrm{B},\mathrm{C},\mathrm{D},\mathrm{A}}/kT)$ are ratio of rates for 
a up and down thermal transitions. For values $\hbar \omega _{\mathrm{B},\mathrm{C},\mathrm{D}}/kT \gg 1$,
$\hbar \omega ^{\pm }_{\mathrm{A}}/kT \ll 1$ $(T \leq 0,1\,\mathrm{K})$ there are the thermal electron $P_{\mathrm{S}0} = - 1$ and
nuclear $P_{\mathrm{I}0} = \hbar \omega _{\mathrm{A}}^{+}/kT \ll 1$ polarizations.\par
Let us propose next that rate $W_{\mathrm{e}}$ of {\it the induced forbidden 
electron transitions }$|1,1\rangle \Rightarrow |1,-1\rangle $ and electron longitudinal 
relaxation times are satisfied the conditions: 
$W_{\mathrm{e}}^{-1} < T_{\parallel \mathrm{B}} \approx T_{\parallel \mathrm{C}}, T_{\parallel \mathrm{A}} \ll T_{\parallel \mathrm{D}} < T_{\parallel \mathrm{S}} (T_{\parallel \mathrm{S}}$ is the longitudinal relaxation 
time of electron spins for forbidden transition). Hereafter we shall 
write\par
\begin{eqnarray}
dp(0,0)/dt = p(1,1)/T_{\parallel \mathrm{B}} + (p(1,-1) - p(0,0))/T_{\parallel \mathrm{A}} ,\nonumber
\end{eqnarray}\begin{eqnarray}
dp(1,-1)/dt = p(1,0)/T_{\parallel \mathrm{B}} + ((p(1,1) - p(1,-1))\cdot W_{\mathrm{e}} +\nonumber
\end{eqnarray}\begin{eqnarray}
+ (p(0,0) - p(1,-1))/T_{\parallel \mathrm{A}} ,\nonumber
\end{eqnarray}\begin{eqnarray}
dp(1,0)/dt = - p(1,0)/T_{\parallel \mathrm{B}} + (p(1,1) - p(1,0))/T_{\parallel \mathrm{A}} +\nonumber
\end{eqnarray}\begin{eqnarray}
dp(1,1)/dt = - p(1,1)/T_{\parallel \mathrm{B}} + (p(1,-1) - p(1,1))\cdot W_{\mathrm{e}} ,\nonumber
\end{eqnarray}\begin{eqnarray}
+ (p(1,0) - p(1,1))/T_{\parallel \mathrm{A}}\label{31}
\end{eqnarray}
With equations (\ref{28}),(\ref{29}),(\ref{31}) we can obtain the rate equations 
for $P_{\mathrm{S}}$ and $P_{\mathrm{I}}$:\par
\begin{eqnarray}
dP_{\mathrm{S}}/dt = - (P_{\mathrm{S}} + P_{\mathrm{I}})\cdot W_{\mathrm{e}} - (P_{\mathrm{S}} + 1)/T_{\parallel \mathrm{B}} ,\nonumber
\end{eqnarray}\begin{eqnarray}
dP_{\mathrm{I}}/dt = - (P_{\mathrm{S}} + P_{\mathrm{I}})\cdot W_{\mathrm{e}} - P_{\mathrm{I}}/T_{\parallel \mathrm{A}}.\label{32}
\end{eqnarray}
The steady-state saturation condition $(W_{\mathrm{e}} \gg 1/T_{\parallel \mathrm{A}})$ of the 
transition $|1,1\rangle \Rightarrow |1,-1\rangle $ gives rise to the equalization of the 
populations $p(1,1) = p(1,-1)$ and respectively to the {\it full nuclear 
spin polarization}\begin{eqnarray}
P_{\mathrm{I}} = - P_{\mathrm{S}} = p(0,0) = 1.\label{33}
\end{eqnarray}
It is obvious that this state is equivalent to state of cooling
up spin temperature $T_{\mathrm{I}} < \hbar \omega _{\mathrm{A}}/k \sim 10^{-3}\,\mathrm{K}$ nuclear spins.\par
Let us estimate now the needed microwave power for saturation.
The rate of external microwave field that induce forbidden electron
transitions with frequency $\omega _{\mathrm{S}}$ (for $X \approx \gamma _{\mathrm{e}}\hbar B/A \leq 1)$ is
\begin{eqnarray}
W_{\mathrm{e}} < (\gamma _{\mathrm{e}}b_{\mathrm{mw}})^{2}\cdot (\overline{\Delta \omega ^{2}_{\mathrm{S}}})^{-1/2} ,\label{35}
\end{eqnarray}
where $(\overline{\Delta \omega ^{2}_{\mathrm{S}}})^{1/2} \approx 2T_{\perp \mathrm{S}}^{*-1}$ --- {\it nonhomogeneous broadened }resonance line
widths for the saturated electron transition, $b_{\mathrm{mw}}$ is amplitude of
{\it microwave field}. The quality factor $Q_{\mathrm{c}}$ of microwave cavity is
\begin{eqnarray}
Q_{\mathrm{c}} \approx \omega _{\mathrm{S}}b^{2}_{\mathrm{mw}}\cdot V_{\mathrm{r}}/(2\mu _{0}P) ,\label{35}
\end{eqnarray}
where $V_{\mathrm{r}}$ --- volume of microwave resonator, $P$ --- dissipating power. The
saturation condition is
\begin{eqnarray}
W_{\mathrm{e}} \gg 1/T_{\parallel \mathrm{S}} \;\mathrm{or}\; (\gamma _{\mathrm{e}}b_{\mathrm{mw}})^{2} T_{\perp \mathrm{S}}^{*} T_{\parallel \mathrm{S}} \gg 1 ,\label{36}
\end{eqnarray}
The power dissipated in cavity is\begin{eqnarray}
P > \omega _{\mathrm{S}}V_{\mathrm{r}}W_{\mathrm{e}}(\overline{\Delta \omega ^{2}_{\mathrm{S}}})^{1/2}/(2\mu _{0}Q_{\mathrm{c}}\gamma _{\mathrm{e}}^{2}).\label{37}
\end{eqnarray}
For example, taking $W_{\mathrm{e}} \sim 10^{3}\,\mathrm{s}^{-1}$, $\omega _{\mathrm{S}} \sim 100$ radGHz, $V_{\mathrm{r}} \sim 1\,\mathrm{cm}^{3}$, 
$Q_{\mathrm{c}} \sim 1000$ and $(\overline{\Delta \omega ^{2}_{\mathrm{S}}})^{1/2} \sim 10^{8}\,\mathrm{s}^{-1}$ \cite{18} as a rough estimate we obtain 
$P > 1\,\mathrm{mW}$. Notice that this power is applied only during the 
saturation process over of time $\geq W_{\mathrm{e}}^{-1} \sim 1\,\mathrm{ms}$ in the act of
initializing of qubit states.\par
Hence, the initialization of nuclear states may be obtained by 
use of ENDOR technique at the lattice temperature of order 0,1 K and 
by this means the {\it the second difficulty }of Kane's scheme can be 
overcame.\par
\par
\section{A silicon-based ensemble NMR quantum computer of cellular-automaton type}
\par
For implementation of ensemble silicon quantum computer, 
operated on cellular-automaton principle, it may be usable the 
previously considered system of long chains of donor atoms $^{31}\mathrm{P}$ 
disposed in silicon, but free of the $\mathbf{A}$ and $\mathbf{J}$ gates.\par
If exchange interaction constant (it is here positive) for 
localized electronic spins of $^{31}\mathrm{P}$ along the chain is more than Zeeman 
energy $J(l) \gg \gamma _{\mathrm{B}}\hbar B$ and the temperature is well below the critical 
temperature for electron ordering (Neel temperature) $T < T_{\mathrm{NS}} \sim J(l)/k$ 
($k = 1,38\cdot 10^{-23}\,\mathrm{J}/\mathrm{K}$ --- the Boltzmann constant), the one-dimensional 
{\it antiferromagnetically ordered ground state }of electronic spins is 
produced. We can estimate the exchange interaction constant as 
$J > \gamma _{\mathrm{N}}\hbar B \sim 6.5\cdot 10^{-23}\,\mathrm{J}$, that correspond to the distance between 
donors $l_{\mathrm{x}} \sim 20\,\mathrm{nm}$ \cite{9}, that is the electron spin critical temperature 
will be $T_{\mathrm{NS}} \sim 4\,\mathrm{K}$.\par
Due to hyperfine interaction nuclear spins will be oriented 
according to the electronic spin direction in the resultant field and 
will form array with the alternating orientation of nuclear spins. At 
magnetic fields $B \leq A/\gamma _{\mathrm{N}}\hbar \sim 3.5\,\mathrm{T}$ and at temperatures $T \sim 10^{-3}\,\mathrm{K}$ the 
nuclear spins $^{31}\mathrm{P}$ will form a periodic ground state array of 
ABAB$\ldots $type: $\uparrow  \downarrow  \uparrow  \downarrow  \ldots $, where $\uparrow $ marks the ground state of nuclear 
spin in an A-site and $\downarrow $ --- the ground state of nuclear spin in a B-site with almost 100\% opposite orientation $(\omega _{\mathrm{A},\mathrm{B}}/kT \geq 1)$. That is the 
distinct nuclear spins are {\it in initializing ground state}.\par
Notice that the using of dynamic methods, makes possible the 
high orientation of nuclear spins also at larger temperatures and, 
according to the above section 5, this state will be {\it the long-lived 
nonequilibrium }nuclear spin state.\par
The nuclear resonant frequencies $\omega _{\mathrm{A},\mathrm{B}}$ of neighbor nuclear spins 
are different for each of the magnetic one-dimensional subarray A and 
B in the chain as they depend on the states of neighboring spins. We 
will take it here in the simple form \cite{12,19}:\begin{eqnarray}
\omega _{\mathrm{A},\mathrm{B}} \equiv \omega (m_{<} + m_{>}) \approx |\gamma _{\mathrm{I}}\hbar B \pm A/2 - I_{\mathrm{n}}(m_{<} + m_{>})|/\hbar ,\label{38}
\end{eqnarray}where $I_{\mathrm{n}} \sim (A^{2}/J)$ is constant of two neighbor nuclear indirect spin-spin interaction due to hyperfine interaction, $m_{<}$ and $m_{>}$ are the 
magnetic quantum numbers for the left and right spin subarray. The 
nonsecular part of nuclear-nuclear interaction is neglected here 
taking into account that $\gamma _{\mathrm{N}}\hbar B$, $A/2 \gg I_{\mathrm{n}}$. The difference between 
nuclear resonant frequencies for the distinct neighboring spins 
orientations is $\Delta \omega _{\mathrm{I}}/2\pi \sim I_{\mathrm{n}}/2\pi \hbar \sim (A^{2}/J)/2\pi \hbar \sim 0.5\,\mathrm{MHz}$, whereas the 
resonant nuclear frequencies are $\omega _{\mathrm{A},\mathrm{B}}/2\pi \sim A/4\pi \hbar \sim 120\,\mathrm{MHz}$.\par
For the organization of logic operations let us use the 
addressing to spin states, analogously to a scheme put forward in
\cite{20}. Each nuclear spin in A-site of this scheme has {\it two internal 
}eigenstates -- ground $|\uparrow \rangle $ and excited $|\Downarrow \rangle $ and in B-site, accordingly, 
-- $|\downarrow \rangle $ and $|\Uparrow \rangle $.\par
We take into account that the life time of excited states (the 
longitudinal nuclear spin relaxation time $T_{\parallel} )$ at low temperatures is 
very long. Each {\it logic qubit }of quantum information in this state will 
be encoded here, similar to \cite{20}, by the {\it four states }of {\it physical 
spin}-{\it qubits }(individual nuclear spin states): the logical qubit basis 
state ''0'' will be encoded by unit $|\Downarrow \Uparrow \uparrow  \downarrow \rangle $, while the state ''1'' --- by 
$|\uparrow  \downarrow  \Downarrow \Uparrow \rangle $. It is important here that the resonant frequencies of 
nuclear spins depend on neighbor spins states. Both logical states 
have two excited spin states and the {\it zero projection }of total nuclear 
spin.\par
Notice that a random inversion of only one spin will result in 
degradation of the qubit state. But to form the rough error, for 
example, of ''0'' $\Rightarrow$ ''1'' type in the coding of stored quantum
information it is essential to invert {\it simultaneously four }spins. 
Therefore, it may be concluded that the considered way of qubit 
coding ensures {\it a better fault-tolerance }with respect to this type of 
errors.\par
As the {\it ports }(noted on Fig. 5 by $\times $) for input and output of the 
information in the array of ground states spin-qubits can be dopant 
nuclei D at the certain place of the array with distinct resonant 
frequency, defects or local gates that modify the resonant frequency 
of the nearest nuclear spin in the array.\par
\par
\begin{center}
\epsfbox{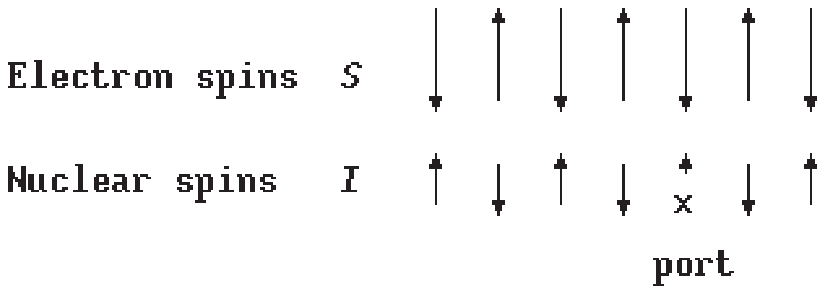}
\nobreak\par\nobreak
Fig. 5. Scheme of electron and nuclear spins ordering\par
\end{center}
$\;$
\par
Starting from the perfectly initialized states inputting the
information can be performed by setting the dopant D-spin to a 
desired state by means of RF-pulse at its resonant frequency. The 
nuclear spin state of spin nearest to the dopant spin is created by 
SWAP operation. After the required information is loaded, D-spin is 
reset to the ground state. Upon completion of computation, the state 
of any spin can be measured by moving it to the A-site nearest to D, 
then swapping A$\Rightarrow \mathrm{D}$ and finally measuring the state of D-spin.\par
For the implementation of quantum operations on logic qubits we 
will also introduce, as in (\ref{19}), one auxiliary {\it control unit }(CU), 
which is represented here by {\it six }physical spin states in the pattern 
$\overline{\overline{\Uparrow \Downarrow \downarrow  \uparrow  \Uparrow \Downarrow }}$. The CU exists only in one place along the array and is 
separated from logical qubits by odd number of spacer spins. The 
applying corresponding SWAP sequence of pulses CU leads to 
interaction of CU with one and two logical qubits and performing on 
they one- and two-qubit quantum operations (for more details, see 
\cite{12,19}).\par
In the case of a large enough ensemble of in parallel acting 
chains the states may be measured by NMR methods. For the increasing 
of logic qubit number in single ''molecule'' to $\geq 10^{3}$, that falls on 
one port, it can be used two- and three-dimensional structures with 
antiferromagnetic chess-type ordering of electron spins. The 
corresponding ordering will have and for nuclear spins 
(Fig.\,6).\par
\par
\begin{center}
\epsfbox{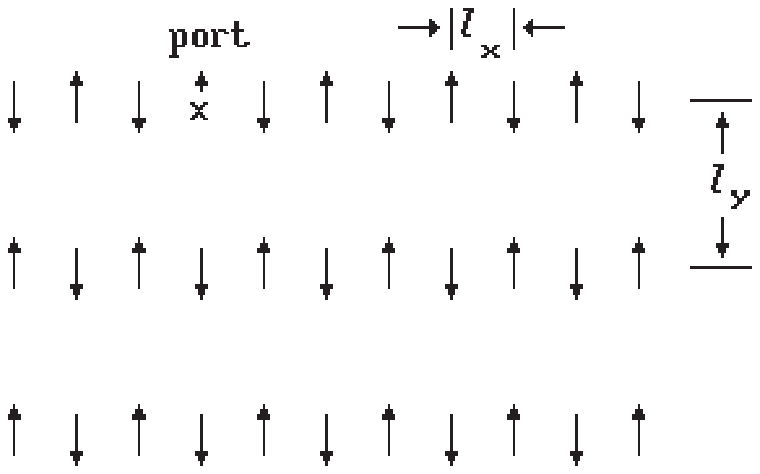}
\nobreak\par\nobreak
Fig. 6. Scheme of two dimensional ordering of initialized
nuclear spins.\par
\end{center}
$\;$
\par
If, for example, the number of spins-qubits, falls on one port 
in linear chain is, say, $L \sim 30$, so in two-dimension case their
number will be $L = 900$. Ensemble that is composed of $N \sim 10^{5}$ in
parallel acting such plane ''artificial molecules'' permit to provide 
input and output of information through the standard NMR techniques. 
The using of more sensitive ENDOR techniques has particular meaning 
when it may be combined with the techniques of dynamic polarization 
(solid state effect).\par
The advantage of ensemble quantum cellular automaton in the 
comparison with the above considered ensemble variant with strip 
gates is as follows:\par
a) The system of the control gates is absent, what essentially 
simplifies the producing of computer structure and eliminates one of 
the important source of decoherence.\par
b) The coding of logic qubits into four physical qubits gives a 
higher degree of fault-tolerance in logic operations.\par
It follows that the going to ensemble quantum cellular automaton 
permits to overcome {\it the third and fourth }difficulty of Kane's scheme.\par
The main disadvantage of cellular automaton scheme is the 
relative complexity of logic operation performing.\par
Notice here that there is also an another possibility of the 
ensemble NMR implementation, which does not have the gate system. The 
selectivity of nuclear resonance frequencies for individual qubits in 
the chain ensemble can be achieved, instead of $\mathbf{A}-$gates voltage, by 
applying the external magnetic field gradients along axis $x$ for 
neighbour qubits separated by $\sim $ 20 nm of order $dB_{\mathrm{z}}/dx \sim 1\,\mathrm{T}/\mathrm{cm}$ (that 
is feasible now), which produce a $\sim 100\,\mathrm{Hz}$ resonance frequency 
difference.\par
\par
\section{Decoherence of the qubit quantum states in an ensemble register}
\par
The interaction of nuclear spins--qubits with polarized {\it random 
distributed} paramagnetic centre (atoms and defects) in substrate 
causes a spread of resonance frequencies for distinct qubits,
determinated by the dipole-dipole electron--nuclear spin
interaction. For low temperatures $(T \leq 0,1\,\mathrm{K})$ it is the main
mechanism of {\it nonhomogeneous broadening }of nuclear resonance line
widths $(\overline{\Delta \omega ^{2}_{\mathrm{I}}})^{1/2}$. Therefore for its elimination it is required to use
the {\it cleaning} from such centres.\par
To estimate the allowed concentration of paramagnetic centres in
silicon substrate $C_{\mathrm{S}}$ we will take the expression
\begin{eqnarray}
(\overline{\Delta \omega ^{2}_{\mathrm{I}}})^{1/2} \sim \mu _{0}/(4\pi )\cdot \gamma _{\mathrm{N}}\gamma _{\mathrm{B}}\hbar C_{\mathrm{S}} \geq 1/T_{\perp \mathrm{I}} \gg 1/T_{\parallel \mathrm{I}} ,\label{39}
\end{eqnarray}
where $T_{\perp \mathrm{I}}$ --- the homogeneous nuclear spin transverse relaxation time,
which correspond to {\it decoherence time} $\tau _{\mathrm{D}}$ of nuclear quantum states.
This time to be {\it sufficiently large} $(\sim 1\,\mathrm{s})$ as compared with the RF
control pulses length $(\sim 10^{-5}\,\mathrm{s})$ in order one can be done in a time 
of order $\tau _{\mathrm{D}}$ a necessary for large-scaled computing number of logic 
operation $(> 10^{5})$ \cite{15}.\par
Taking into account that frequency of interqubit indirect 
interaction $\nu _{\mathrm{J}} \approx 100\,\mathrm{kHz}$ (see above), for the {\it coherence length }of 
quantum register state we obtain $\lambda \sim 2\pi \Delta \nu _{\mathrm{I}}l_{\mathrm{x}}\cdot \tau _{\mathrm{D}} \approx 6.3\cdot 10^{5}\cdot l_{\mathrm{x}}$.\par
With $\tau _{\mathrm{D}} = 1$ we obtain for the allowed concentration of
paramagnetic centers the estimate $C_{\mathrm{S}} < 10^{15}\,\mathrm{cm}^{-3}$, that is {\it almost usual
}requirement for {\it a degree of cleaning }from unwanted impurities in 
silicon technology.\par
The another mechanism of decoherence of qubit states is the 
dipole-dipole interaction with {\it nuclear spins }of impurity atoms of 
nuclear spin containing elements.\par
The {\it allowed concentration C}$_{\mathrm{N}}$ of impurity with nuclear spins in 
silicon substrate can be estimated from the following expression\par
\begin{eqnarray}
\tau _{\mathrm{D}} > ((\mu _{0}/4\pi )\cdot \gamma _{\mathrm{N}}^{2}\hbar \cdot C_{\mathrm{N}}.\label{40}
\end{eqnarray}
Taking again $\tau _{\mathrm{D}} \sim 1\,\mathrm{s}$, we will obtain $C_{\mathrm{N}} < 2\cdot 10^{18}\,\mathrm{cm}^{-3}$. For the 
number of silicon atom $5.0 \cdot 10^{22}\,\mathrm{cm}^{-3}$ the allowed relative 
concentration of the impurity nuclear spin, contained isotope $^{29}\mathrm{Si}$,
must be $C_{\mathrm{N}}\% < (2\cdot 10^{18}\cdot 100)/(5.0\cdot 10^{22}) = 0.4\cdot 10^{-2}\%$. For comparison,
the natural abundance of isotope $^{29}\mathrm{Si}$ in natural silicon is 4.7\%. At
present the realized degree of cleaning $^{28}\mathrm{Si}$ is 99,98\%, which is yet
{\it not sufficient }to our purposes.\par
The allowed concentration for another impurity elements having
the natural abundance of isotopes with nuclear spins $P < 100\%$ is 
determined from\begin{eqnarray}
C_{\mathrm{N}}\% < 10^{-2} 100/P.\label{41}
\end{eqnarray}
In the case of ensemble approach for implementing of logic 
quantum operations, for read-out of information and for suppressing 
of decoherence effects, caused by secular part of dipole-dipole 
interaction qubits with randomly distributed paramagnetic centres, 
may be explored the well known {\it many-pulsed technique }of {\it high 
resolution }NMR spectroscopy in solids \cite{21}.\par
As source of errors in quantum computer operation there is also 
generation processes unwanted states for neighbour initialized qubits 
as a result of nonresonance effects. This is a further mechanism of 
the decoherence of quantum states. The probability of such errors can 
suppress to the required value $(< 10^{-5})$ by choosing of Rabi frequency 
value for control RF-pulses \cite{22}. For one example, if resonance 
frequencies of nonresonant qubits differ from the frequency of 
resonant qubit by $\Delta \omega $, so the Rabi frequency for $\pi -$pulse should be 
$\Omega = |\Delta \omega |/\sqrt{4k^{2}- 1}$, where $k = 1, 2, 3, \ldots $\par
Notice that it was also suggested another, as compared to 
considered above, scheme of the four-spin encoding two logical 
qubits, which are represented by the two zero-total spin states of 
four spins, generated by the pairs respectively of the singlet and 
triplet states \cite{23}. This scheme is based on the so named {\it collective 
decoherence conditions}. This conditions can be attained in coupled 
spins at very low temperatures, when the longest wavelength magnon 
modes are quenched \cite{12,19}.\par
The {\it chief disadvantage }of the variant with strip gates as and 
Kane's scheme is associated with availability of the gates. The 
thermal voltage fluctuations on gates are additional mechanism of 
quantum state decoherence. They are essentially suppressed here if 
both the Al gate electrodes and the entire resonance circuit are 
superconductors at the work temperature $T < T_{\mathrm{C}}$ ($T_{\mathrm{C}} = 5.8\,\mathrm{K}$ for 
films).\par
\par
\section*{Conclusion}
\par
\par
1) The line of the large-scale ensemble NMR quantum computer 
development have certain advantage over the line, based on Kane's 
scheme. It consists in the possibility of employment of the standard 
NMR technique for the measurement of quantum states at output of 
computer, like in the liquid prototype.\par
2) For the initialization of nuclear spin states at temperature 
$T \sim 0,1\,\mathrm{K}$ methods of dynamic polarization of solid state effect type 
may be proposed.\par
3) Analysis of proposed planar structure of ensemble silicon 
computer shows the possibility of realization of large-scale NMR 
quantum computer for ensemble component number $N \sim 10^{5}$.\par
4) The implementation of cellular automaton principle permits to 
abandon the realization regular nanostructure in form of gate chains.\par
5) Analysis of different feasible ways for to obtain a 
decoherence times large enough shows that the values, needed to 
perform required for large-scaled computations number of quantum 
logic operations $\sim 10^{5}$, can be achieved.\par
6) One would expect, that in the future the combined variants of 
solid state quantum computers will be proposed. They can, say, 
exploit the structures that are contained nuclear spins and quantum 
dots with electron spins and the combined methods access to qubits, 
like ENDOR method.\par
\par
\section*{Appendix}
\subsection*{Signal NMR for discrete ensemble of nuclear spins}
\par
Let us consider here the sample that involves $N$ = $nL\cdot pN_{0}$ nuclear 
spins arranged in the plane $z = 0$ of the silicon plate at regular 
intervals along the strips (Fig. 4). The spins in chains under strip 
at resonance in the each block are oriented along $x$ axis (solenoid 
axis) and separated at intervals of $L$. The read out NMR signal is
\begin{eqnarray}
|V_{\mathrm{max}}| = Q \omega _{\mathrm{A}} \frac{K}{X} \int_{-X/2}^{X/2} |\int_{A} B_{\mathrm{xmax}}(x,y,z) {\it dydz}| dx ,\label{A.1}
\end{eqnarray}
where $X = nL \gg L$ is roughly the length of solenoid and\par
\begin{eqnarray}
B_{\mathrm{xmax}}(x,y,z) =\label{A.2}
\end{eqnarray}
\begin{eqnarray}
= \frac{\mu _{0}\gamma _{\mathrm{I}}\hbar }{16\pi } {\sum_{n_{\mathrm{i}}=-n/2}^{n/2}} {\sum_{p_{\mathrm{i}}=-pN_{0}/}^{pN_{0}/2}}_{2} \frac{- 2(x - Ln_{\mathrm{i}})^{2} + (y - l_{\mathrm{y}}p_{\mathrm{i}})^{2} + z^{2}}{[(x - Ln_{\mathrm{i}})^{2} + (y - l_{\mathrm{y}}p_{\mathrm{i}})^{2} + z^{2}]^{5/2}}\label{A.3}
\end{eqnarray}
is the peak magnetic field produced by resonant spins in solenoid. 
For simplicity it is suggested that $n, p, N_{0}, L \gg 1$ are the even 
numbers. We assume that the area of coil turns is A = $D\cdot \delta $ 
 ($D = l_{\mathrm{y}}\cdot pN_{0}$, $\delta \ll D$).\par
For summation over $n_{\mathrm{i}}$ and $p_{\mathrm{i}}$ we have used the {\it Poisson's 
summation formula}, namely,\par
\begin{eqnarray}
{\sum_{p_{\mathrm{i}}=-pN_{0}/2}^{pN_{0}/2}} f(l_{\mathrm{y}}p_{\mathrm{i}}) = \frac{pN_{0}}{D} {\sum_{\nu =-\infty }^{\infty }} \int_{-D/2}^{D/2} f(\xi ) \exp (i\nu 2\pi \xi /l_{\mathrm{y}}) d\xi ,\label{A.4}
\end{eqnarray}
by omitting the oscillated terms with $\nu \neq 0$:\par
\begin{eqnarray}
|V_{\mathrm{max}}| = \mu _{0} QK \omega _{\mathrm{A}} \frac{\gamma _{\mathrm{I}}\hbar }{8\pi } \frac{1}{X} \int_{-X/2}^{X/2}dx \frac{n}{X} \int_{-X/2}^{X/2}d\eta \cdot \nonumber
\end{eqnarray}\begin{eqnarray}
\cdot \int_{0}^{\delta /2} dz |{ \frac{pN_{0}}{D} \int_{-D/2}^{D/2} \int_{-D/2}^{D/2} \frac{- 2(x - \eta )^{2} + (y - \xi )^{2}+ z^{2}}{[(x - \eta )^{2}+ (y - \xi )^{2}+ z^{2}]^{5/2}} dyd\xi }| \label{A.5}
\end{eqnarray}
Taking into account $D^{2} \gg z^{2}$, upon integrating (A.4) over $y$, $\xi$, we obtain
\begin{eqnarray}
|V_{\mathrm{max}}| = \mu _{0}QK \omega _{\mathrm{A}} \frac{\gamma _{\mathrm{I}}\hbar }{8\pi } \frac{pN_{0}}{D} \frac{1}{X} \int_{-X/2}^{X/2} dx \frac{n}{X} \int_{-X/2}^{X/2}d\eta \cdot \nonumber
\end{eqnarray}
\begin{eqnarray}
\cdot \int_{0}^{\delta /2} dz \left\{{ \frac{2[(x-\eta )^{2}-z^{2}] D^{2}}{[(x-\eta )^{2}+ z^{2}]^{2} [D^{2}+(x-\eta )^{2}]^{1/2}} + \frac{2 z^{2}}{[(x-\eta )^{2}+ \mathrm{z}^{2}]^{3/2}} }\right\}.\label{A.6}
\end{eqnarray}
By integrating now over $z$ and preserving only the major
logarithmically increasing for 2$|x - \eta |/\delta \Rightarrow 0$ term, we obtain\begin{eqnarray}
|V_{\mathrm{max}}| \approx \mu _{0}{\it QKA }\omega _{\mathrm{A}} \frac{\gamma _{\mathrm{I}}\hbar }{4\pi } \frac{npN_{0}}{AD} \frac{1}{X} \int_{-X/2}^{X/2} dx \frac{1}{X} \int_{-X/2}^{X/2} d\eta \log \frac{\delta }{|x - \eta |} =\nonumber
\end{eqnarray}\begin{eqnarray}
= (\mu _{0}/4) {\it QKA }\omega _{\mathrm{A}} \frac{N}{V_{\mathrm{s}}} \gamma _{\mathrm{I}}\hbar \cdot \frac{X}{\pi D} \log \frac{X}{\delta \sqrt{\mathrm{e}}} .,\label{A.7}
\end{eqnarray}
where $V_{\mathrm{s}} = AX$, $N = npN_{0}$, e = 2,718\ldots .\par
We see that expression (\ref{A.7}) is distinguished from (\ref{4}) by non-essential factor $\frac{X}{\pi D}\,\log \frac{X}{\delta \sqrt{\mathrm{e}}}$, that is of order several ones.\par
\par
\par

\end{document}